\documentclass[preprint,aps]{revtex4}

\usepackage{graphicx}
\usepackage{dcolumn}
\usepackage{bm}
\usepackage[usenames]{color}    

\usepackage{graphicx}
\usepackage{bm}
\usepackage{color}

\newcommand{\be}{\begin{equation}}
\newcommand{\ee}{\end{equation}}
\newcommand{\bea}{\vspace{0.25cm}\begin{eqnarray}}
\newcommand{\eea}{\end{eqnarray}}

\def\PRL{{\it Phys. Rev. Lett.} }
\def\PRA{{\it Phys. Rev.} A }

\begin{document}

\title{ Can quantum non-locality be connected to extra-dimensions?}

\author{Marco Genovese }

\address{ Italian National Metrology Research Institute (INRIM), Strada delle Cacce 91, I-10135 Turin, Italy and INFN, sede di Torino, via P.Giuria 1, 10125 Turin, Italy. }

\begin{abstract}
Quantum non locality, as described by EPR paradox, represents one of the mysteries at the very foundations of quantum mechanics. Here we suggest to investigate if it can be understood by considering extra dimensions.
\end{abstract}
\maketitle

\section{Introduction}

Quantum non-locality,  is a fundamental argument of the debate on the foundations of quantum mechanics (QM)\cite{bell,mm,cl71,as82,lf1,lf2,lf3,int,eli18,gis20,gra21}. Furthermore, it is also becoming an important resource for emerging quantum technologies \cite{qt1,qt2,qt3,qt4,qt5,qt6}. This prompts the need of a clear understanding of quantum non-locality and, in particular, of its compatibility with special relativity.

Even if one can rigorously demonstrate that it can not lead to any superluminal transmission (signaling) \cite{GRW80}, peaceful coexistence between special relativity and QM would require more, i.e., it would be necessary to understand a coherent description among different observers \cite{shi}. For instance, how to reconciliate two observers that see a different temporal order in the collapse of two entangled particles?  The answers to this question span from being an unsolved problem \cite{mau,Macc19,st,st2,st3,st4,st5,st6}, requiring a preferred foliation to relativistic space--time, to being only apparent, since accounts of entangled systems undergoing collapse yielded by different reference frames can be considered as no more than differing accounts of the same process and events \cite{Myr};  there is a form of holism associated to QM description of composite systems \cite{dar21}, the factorizable state after the collapse on a certain hypersurface is merely one way of slicing together local parts, entangled states are a superposition of such splicings. Also a connection between quantum non-locality and wormholes, summarized by the sentence "ER=EPR" (Einstein Rosen wormholes are tied to Einstein-Podolsky-Rosen paradox) has been suggested by Maldacena and Susskind \cite{ER} or the question looses its problematic aspects in superdeterministic theories \cite{th1,t1}.

\section{A solution in higher dimensional spaces}

 Be that as it may, rebus sic stantibus, an undoubted difficulty in understanding the collapse in composite systems persists.
 This is true for standard quantum mechanics, but it is even more important for modifications of quantum mechanics formalism where a spontaneous collapse occurs. For instance, Ref.s \cite{c1,c2} demonstrated that creating a relativistic version of spontaneous collapse models "à la GRW" is impossible, while Ref.s \cite{c,g,p} showed that certain collapse models lead to superluminal signaling. Furthermore while Bell inequalities tests exclude local hidden variable theories \cite{bell,mm,cl71,as82,lf1,lf2,lf3,int}, non-local hidden variable theories still remain a valid alternative to standard quantum mechanics \cite{int,db1,db2,q1,q2,q3}, but in this case non-locality must be understood in their framework.

Here we suggest that one should explore the possibility that the  quantum non-locality "problem" can be solved in higher dimensional spaces \cite{mm}.
In the following we will present in little more detail this idea suggested in \cite{mm}, discussing some motivations and detailing some difficulties to be considered in developing it. It is not our intention to provide a precise model: probably several different ways can be gone along leading to rather different models and various mathematical problems must be solved on these paths. We limit ourselves to introducing this new idea, hoping this letter can stimulate further studies contributing significantly to the debate on quantum non-locality.

 Let us consider a $n>4$ space-time with $n-4$ more (spatial) dimensions and let us suppose that, while the usual fields only ``live'' in $3+1$ dimensions, the collapse involves also other dimensions, eventually being induced by ``some field'' propagating also in these extra-dimensions. We are considering the collapse on the position variable, since in general every measurement at the end reduces to a position measurement.

 Every n-dimensional space can be ``contracted'' around a single point of a higher dimensional space. For instance, a plane can be wrapped as tied as one  wishes around a single point in three-dimensionall space:  two points of a napkin can be several centimeters apart moving on the napkin, but they can be very next in the third dimension after squeezing the napkin in a hand.  Thus, two far points in the ``ordinary'' space can be as close as wanted considering extra-dimensions. Incidentally, this ``contraction'' is an isometry (the distances in the 3+1 are unchanged)  and leaves invariant the Gaussian curvature. Nonetheless, a precise characterisation of the effects on local curvature would depend on the specific model and would represent a significant point in elaborating specific models.

If this extra-dimensions are extremely small (for example around the Planck scale, as supposed for instance by some quantum gravity theory \cite{qg,m} ), every event of the 3+1 dimensional space could be connected with any other event in an ``extremely short'' effective distance through the compactified dimensions. Therefore, the collapse of one of two remote entangled particles could cause the collapse of the other with the propagation of a subluminal signal through the compactified dimensions. Quantum non-locality would reduce to the fact that only the wave function collapse would be affected by these extra dimensions.

 This kind of phenomenon could either be included in a more general theory going beyond quantum mechanics (for example, some Planck scale theory already predicts compactified dimensions) or to find a use in building relativistic collapse models \cite{r,r1,r2,r3,r4,r5}. For example a simple generalization of Tumulka's one \cite{tum} should eventually be able to incorporate it.

As a simple example, one could eventually consider some general non-unitary evolution operator
\be U(t) = Exp \{ - i \int d^4 x d^n y {\it O}_I(x,y,t)\} \ee
where $ {\it O}_I(x,y,t) = {\it H}_I(x,t) + C_I(x,y,t)$, ${\it H}_I(x,t)$ being the usual Hermitian--Hamiltonian density of  quantum fields leaving in a 3+1 dimensional space (x), while the component $C_I(x,y,t)$ also inclues the extra dimensions, y. $C_I(x,y,t)$ is not Hermitian and eventually induces the collapse by inducing a non-unitary evolution. When one traces over the degrees of freedom corresponding to the not Hermitian component, $C_I(x,y,t)$, the usual unitary evolution is eventually restored, as in 't Hooft finite degrees of freedom model \cite{th}.

For instance, let's suppose one has a scalar field with a usual Hamiltonian, where x are usual space time coordinates and y additional compactified dimensions
$$
1/2  \partial \Psi^*(x,y) \partial \Psi (x,y) +  g(x,y) (\Psi^*(x,y)\Psi (x,y))^n
$$
if the coupling g has a real part depending on x and an imaginary part depending on y the total Hamiltonian is not Hermitian and originates a non-unitary evolution.
Nonetheless, when tracing over the y degrees of freedom one recovers a Hermitian Hamiltonian and a unitary evolution.
In particular one can consider the case when the evolution operator, containing the non Hermitian component of the Hamiltonian, reduces to a projector collapsing the state in a precise position with a probability of collapse proportional to the number of matter field particles ($ b  ^\dagger b$). This guarantees, as in all collapse models, the persistency of quantum superpositions for microscopic objects and a smooth transition to localised objects for macroscopic ones:
\be
U = Exp[- \int  d^dy C_I(x,y,t)] \approx Exp[- g (x-q)^2 b^\dagger(x) b(x)]
\ee
where $C_I(x,y,t)$ can eventually be evaluated by inverting Eq.2.

In alternative the collapse induced by these effects can be included in a Lindblad equation on the line of \cite{lin}.

An interesting point becomes the causal connection among events. Actually, two points in space can be very far in the usual dimensions and therefore two events in these space locations cannot be connected by a causal signal if not after a long propagation. Nevertheless, they are rapidly connected when the collapse is considered, since this propagates in the compactified extra dimension.If the effect of the signal through extra dimensions only affects the collapse, this would not allow faster than light signaling in usual dimensions. For example, Ghirardi-Rimini-Weber theorem \cite{GRW80} guarantees that one cannot use EPR correlations and the collapse induced by one observer for superluminal signaling.

Another consideration concerns the recent theorem demonstrated by Ref. \cite{b22} that collapse dynamics must always be diffusive: the hypotheses of this theorem (in particular the one related to translational invariance) could not be necessary in the models discussed here.

In summary, while one can suppose that usual quantum evolution/interactions  only happen in ordinary space with a unitary evolutions, the collapse of the wave function could be related to the non-unitary evolution in additional dimensions, that being compactified allows a "practically" instantaneous collapse independently by distance in ordinary space. This somehow reproduces t'Hooft model \cite{th} in an infinite dimensional field theory.

\section{Conclusions}

In conclusion, we suggest that quantum non-locality and the problems connected to the temporal order of wave function collapse for entangled systems can eventually find a solution by considering extra dimensions.

We do not propose a specific mathematical model, that can be connected, for instance, to an eventual theory describing quantum gravity, but we suggest that this idea is worth of being further explored.

 Finally, it is worth considering the question of whether such a scheme could lead to observable physical consequences. Of course, models introducing spontaneous collapse of the wave function though this mechanism would lead to the same effects expected in general in collapse models \cite{ab,ab1,ab2,11}.

  Furthermore, one could expect these extra dimensions (and eventual extra field/fields) could provide several possible observable effects. Long-distance effects deriving from crossing these extra dimensions can eventually appear, for example inducing correlations otherwise forbidden by propagation in ordinary dimensions. Just by speculating a little these correlations could eventually even be the source of ultra-quantum technologies or affect present quantum technologies (as a threat to QKD security proofs).
Furthermore, vacuum energy (eventually in connection to cosmologic dark energy) can be affected by this kind of phenomenon.
Nonetheless, the study of different possible models in this general framework and their possible effects is beyond the purpose of this paper simply introducing this idea.


\section{Acknowledgements}

This research was supported by San Paolo Fundation, project QuaFuPhy.

\end{document}